\documentclass[twocolumn]{article}
\usepackage[utf8]{inputenc}
\usepackage[margin=17mm]{geometry}
\usepackage{fancyhdr}
\usepackage{amsmath}
\usepackage{amsfonts}
\usepackage{amssymb}
\usepackage{graphicx}
\usepackage{verbatim}
\usepackage{siunitx} 
\DeclareSIUnit{\mas}{mas}
\DeclareSIUnit{\arcsec}{as}
\usepackage{acronym} 

\usepackage[long,nodayofweek]{datetime}

\usepackage{authblk}
\usepackage{hyperref}

\providecommand{\keywords}[1]
{
  \small	
  \textbf{\textit{Keywords---}} #1
}
\title{Equivalence of Active and Passive Gravitational Mass Tested with Lunar Laser Ranging}
\author[1]{Vishwa Vijay Singh\footnote{Corresponding author:\newline Vishwa Vijay Singh, email: \href{mailto:singh@ife.uni-hannover.de}{singh@ife.uni-hannover.de}}}
\author[1]{Jürgen M\"uller}
\author[1]{Liliane Biskupek}
\author[2]{Eva Hackmann}
\author[2]{Claus L\"ammerzahl}
\affil[1]{\small Institute of Geodesy (IfE),\ Leibniz University Hannover,\ Schneiderberg 50,\ 30167 Hannover,\ Germany}
\affil[2]{\small Center of Applied Space Technology and Microgravity (ZARM), University of Bremen, Am Fallturm, 28359 Bremen, Germany}

\usepackage{framed}
\usepackage{csquotes}
\usepackage{enumitem}
\usepackage[para,online,flushleft]{threeparttable}
\usepackage{breakcites}
\usepackage{multirow}
\usepackage{natbib}
\setcitestyle{square}
\usepackage{subcaption} 
\usepackage{tabularx}

\usepackage[para,online,flushleft]{threeparttable}
\newcolumntype{P}[1]{>{\centreing\arraybackslash}p{#1}} 

\usepackage{url}
\usepackage[normalem]{ulem}
\begin{document}

\maketitle
\thispagestyle{fancy}




\begin{abstract}
\ac{LLR} measures the distance between observatories on Earth and retro-reflectors on Moon since 1969. In this paper, we study the possible violation of the equality of passive and active gravitational mass ($m_{a}/m_{p}$), for Aluminium (Al) and Iron (Fe), using \ac{LLR} data. Our new limit of $3.9\cdot10^{-14}$ is about 100 times better than that of \citet{bartlett_vanBuren_1986} reflecting the benefit of the many years of \ac{LLR} data. 
\end{abstract}



\section{Introduction}\label{sec:intro}

In a non-relativistic framework each body has three masses: the inertial mass, the passive gravitational mass (weight) reacting on a given gravitational field, and the active gravitational mass creating a gravitational field. In standard physics all three masses are assumed to be the same. However, if they are different \citep{Bondi_1957} then for any two gravitationally bound bodies $A$ and $B$ the equations of motion read
\begin{eqnarray}
m_{\text{i}A} \ddot{\boldsymbol{x}}_A & = & m_{\text{p}A} G m_{\text{a}B} \frac{{\boldsymbol{x}}_B - {\boldsymbol{x}}_A}{|{\boldsymbol{x}}_B - {\boldsymbol{x}}_A|^3} \\
m_{\text{i}B} \ddot{\boldsymbol{x}}_B & = & m_{\text{p}B} G m_{\text{a}A} \frac{{\boldsymbol{x}}_A - {\boldsymbol{x}}_B}{|{\boldsymbol{x}}_A - {\boldsymbol{x}}_B|^3} \, ,
\end{eqnarray}
where $m_{\text{i}A}$, $m_{\text{a}A}$, $m_{\text{p}A}$ are the inertial, active and passive gravitational mass of body $A$, respectively. We define relative and center of mass coordinates according to
\begin{eqnarray}
\boldsymbol{x} & = & {\boldsymbol{x}}_B - {\boldsymbol{x}}_A \\
\boldsymbol{X} & = & \frac{m_{\text{i}A}}{M_{\text{i}}} {\boldsymbol{x}}_A + \frac{m_{\text{i}B}}{M_{\text{i}}} {\boldsymbol{x}}_B
\end{eqnarray}
with the total inertial mass $M_{\text{i}} = m_{\text{i}A} + m_{\text{i}B}$. While the relative coordinate evolves according to the Kepler problem
\begin{equation}
\ddot{\boldsymbol{x}} = - G \alpha \frac{\boldsymbol{x}}{x^3} \; , \quad \alpha = \frac{m_{\text{p}A}}{m_{\text{i}A}} m_{\text{a}B} + \frac{m_{\text{p}B}}{m_{\text{i}B}} m_{\text{a}A} \label{xKepler}
\end{equation}
the center of mass coordinate shows a self acceleration
\begin{equation}
\ddot{\boldsymbol{X}} = G \frac{m_{\text{p}A} m_{\text{p}B}}{M_{\text{i}}} S_{A,B} \frac{\boldsymbol{x}}{x^3}
\end{equation}
where
\begin{equation}
S_{A,B} = \frac{m_{\text{a}B}}{m_{\text{p}B}} - \frac{m_{\text{a}A}}{m_{\text{p}A}}  \label{eq:coeff_S}
\end{equation}
describes the difference of the ratio of active and passive masses of the two bodies. $\boldsymbol{x}$ describes the vector between the two bodies. It is interesting that the relative motion decouples from the center of mass motion so that for the determination of the latter any relative motion can be taken. Accordingly, in the case of binary systems this is given by the solutions of equation \eqref{xKepler}, in our case it is the vector between two components of the Moon (see below). This leads to a change in the Earth-Moon distance what can be very precisely measured using \ac{LLR}. In fact, many aspects of gravity have been tested with LLR with the best precision \citep{Muller_etal_1996,hofmann_mueller_2018}.  

From the space mission MICROSCOPE the equality of inertial and passive gravitational mass has been confirmed at the level of $10^{-15}$ in the E\"otv\"os coefficient \citep{MICROSCOPE:2022doy}. From laboratory experiments the equality of active and passive gravitational mass has been shown at the level of $10^{-5}$ \citep{Kreuzer_1968}. Later, \citet{bartlett_vanBuren_1986} used Lunar Laser Ranging (LLR) to improve this estimate to the level $\leq 4 \cdot 10^{-12}$. This limit has further been used to determine other limits, such as on the parameterized post-Newtonian (PPN) parameter $\zeta_3$ \citep{clifford14}. Here we will report on a further improvement on the limit by two more orders of magnitude using a longer and improved set of LLR data. 

\citet{bartlett_vanBuren_1986} considered the self force $F_{\text{s}} = M_{\text{i}} \ddot{X} = S_{A,B} G m_{\text{p}A} m_{\text{p}B}/r_{AB}^2$ between the different parts of the Moon (with the simplifying assumption that the mantle has the same composition as maria, i.e. Iron (Fe) rich basalt, and the crust has the same composition as the highlands, i.e. Aluminium (Al) rich anorthosite). The effect of this force in the tangential direction with respect to the Earth will lead to an increase in the angular velocity of the Moon. Using Kepler's law, stating that $\omega^2r^3$ stays constant, and the change in energy per lunar sidereal month, they express the relation between the self force and the angular velocity of the Moon as,
\begin{equation}\label{eq:om_S}
\frac{\Delta\omega}{\omega} = 6\pi \frac{F_{\text{t}}}{F_{\text{EM}}} \, ,
\end{equation}
where $F_{\text{EM}}$ is the gravitational force between Earth and Moon, $F_{\text{t}} = F_{\text{s}} \sin\delta_{\text{c.m.},\text{c.f.}}$ is the tangential part of the self force using the offset angle $\delta_{\text{c.m.},\text{c.f.}}$ between the directions of the center of mass (c.m.) and the centre of figure (c.f.) of the Moon.

\citet{bartlett_vanBuren_1986} use the offset between the c.m. and c.f. of the Moon, as given by \citet{bills_ferrari_77}, of 1.98 $\pm$ \SI{0.06}{\kilo\meter} in the direction 14.00 $\pm$ \SI{1.00}{\degree} to the east of the vector pointing to the Earth. They also assume a two component Moon, with the mantle having a density of \SI{3.35}{\gram/\centi\meter^3} and the crust having a density of \SI{2.90}{\gram/\centi\meter^3}. Using these assumptions, they show a ratio between $F_{\text{s}}$ and $F_{\text{EM}}$ of
\begin{equation}\label{eq:S_F_ratio}
\frac{F_{\text{s}}}{F_{\text{EM}}} \approx 5 S_{A,B} \, ,
\end{equation}

and give $S_{\text{Al},\text{Fe}} = S_{A,B}/0.08$.

For the uncertainty of the value of the tidal acceleration in the orbital mean longitude of the Moon, $\dot\omega$, \citet{bartlett_vanBuren_1986} considered two values of $\dot\omega$, -25.30 $\pm$ 1.20 arcsec/century$^{2}$ \citep{dickey_et_al1984} and -25.50 $\pm$ 1 arcsec/century$^{2}$ \citep{Christodoulidis_etal_1988}. Using the maximum difference between these values of about 2 arcsec/century$^{2}$, together with equations (\ref{eq:om_S}) and (\ref{eq:S_F_ratio}), they derived an upper limit on the coefficient $S_{\text{Al},\text{Fe}}$ of $4\cdot10^{-12}$.

In the present study we use the latest results from \acf{LLR} to determine a new limit on the coefficient $S_{\text{Al},\text{Fe}}$ by using the current value of $\Delta\omega$. For the study, the ephemeris calculation model is primarily based on the DE430 model \citep{jplde} with only a few minor changes. We consider fourteen solar system bodies (Sun, Moon, eight planets, Ceres, Pallas, and Vesta) for our ephemeris calculation, and use their initial positions at J2000 from the DE430 ephemeris. The initial position and orientation of the Moon (for orientation, mantle and core of the Moon) is fitted to \ac{LLR} data along with other parameters such as station and reflector coordinates, spherical harmonic coefficients of the Moon, etc. A full list of the parameters fitted in a standard calculation can be found in \citet{singh_etal_2021}. For this study, \num{30172} normal points (NPs) in the time span April 1970 - April 2022 were used, obtained thanks to the continuous efforts of various personnel at the different observatories (see Acknowledgements).

\section{Determination of the Lunar Angular Acceleration}\label{sec:meth}
\begin{figure}[ht]
    \centering
    \includegraphics[width=0.49\textwidth]{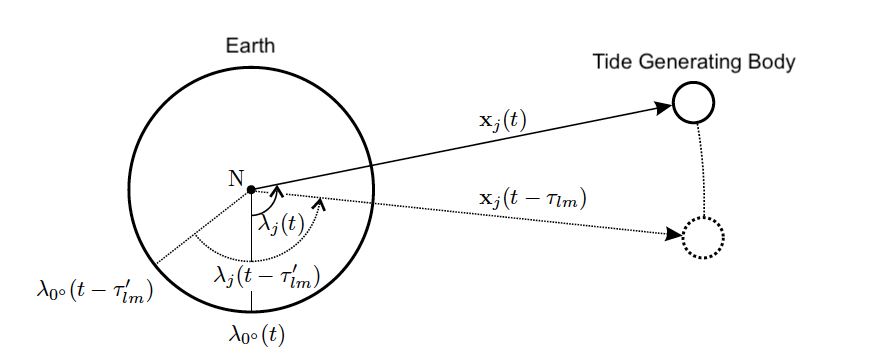}
    \caption{Geocentric change of position of the tide generating body due to time delay \citep{hofmann17}.}
	\label{fig:llr_tau}
\end{figure}

The effect on the lunar orbit of the degree 2 tides on Earth is modelled as a tidal bulge (see Fig. \ref{fig:llr_tau}). This tidal bulge can be modelled, in a simplified way, as one angle defining a geometric rotation \citep{williams_etal_1978}. It used one Love number defined for degree 2. An expanded model has a more complex definition, involving three Love numbers for each frequency of degree 2, and five tidal time delays (three orbital and two rotational) which define the time-delayed position of the tide generating bodies \citep{jplde,williams_boggs_2016}. In this study, we use both methods to obtain a value of $\dot\omega$ and its uncertainty, and compare them to $-25.82 \pm 0.03\; \text{arcsec/century}^{2}$ \citep{jplde}.

For the calculation based on the one angle defining a geometric rotation \citep{williams_etal_1978}, we used an initial value of $k_2\delta = 0.01220$. For the calculation of the ephemeris, all aspects (initial values of solar system bodies, constants, acceleration models, etc.), except the calculation of the effect on the lunar orbit of the degree 2 Earth tides, are based on the DE430 ephemeris \citep{jplde}. All standard parameters except the two rotational components of degree 2 time delays (which were not used in the calculation) were estimated using a \ac{GMM}. We obtained a value of $k_2\delta = 0.01312 \pm 1.17 \cdot10^{-6}$. Using $\dot\omega = -1961\;k_2\delta$ \citep{williams_boggs_2016}, we get $\dot\omega = -25.73 \pm 0.0023\;\text{arcsec/century}^{2}$. This solution is later referred to as \enquote*{k2d}.

For a second solution, later referred to as \enquote*{LUNAR}, we added the effect of degree 2 Earth tides in the \ac{LLR} analysis based on \citet{jplde}. We created four different cases. For three of these cases, along with other standard parameters, the individual values of the three orbital time delays were adjusted. For the fourth case, the three orbital tidal time delay values were kept fixed, and the two rotational tidal time delay values were adjusted along with all other standard parameters. We obtain four sets of values of the five tidal time delays from these variations, which are converted [Williams, 2022; personal communication] to determine four values of $\dot\omega$: -25.7898, -25.7759, -25.7635, and -25.7649 arcsec/century$^{2}$. The uncertainty of $\dot\omega$, taken as the range of the four individual cases, is then obtained as $\pm 0.0263\;\text{arcsec/century}^{2}$.

\section{Discussion}
Different values of $\Delta\omega/\omega$ can be obtained from the different values of the uncertainty mentioned in section \ref{sec:meth}. We use the values of all constants and apply the same assumptions as that used by \citet{bartlett_vanBuren_1986} to recalculate a limit on the violation of the equality of passive and active gravitational mass for Al and Fe. This is done to be able to assess the contribution of the many years of \ac{LLR} data in establishing the limit. The assumptions, such as the \SI{14}{\degree} offset angle between the c.m. and the c.f. of the Moon, an onion-skin lunar interior, $\dot G$ = 0, etc., are critical to the results. Any change compared to these assumptions would affect the results as well. The updated $\Delta\omega$/$\omega$ values are given in Table \ref{tab:res}, along with the limit on the coefficient $S_{\text{Al},\text{Fe}}$ for both solutions (see equations (\ref{eq:om_S}) and (\ref{eq:S_F_ratio})).

\begin{table}[ht]\centering
\caption{Limit on the violation of the equality of passive and active gravitational mass for Al and Fe from the value of $\Delta\omega$/$\omega$ obtained based on our k2d and LUNAR solutions.}\label{tab:res}
\begin{tabular}{ccc}
\hline\noalign{\smallskip}
Solution & $\Delta\omega$/$\omega$ [month$^{-1}$] & $S_{\text{Al},\text{Fe}}$ \\
\hline\hline\noalign{\smallskip}
k2d & $1.2\cdot10^{-15}$ & $6.9\cdot10^{-16}$\\
\noalign{\smallskip}\hline\noalign{\smallskip}
LUNAR & $1.4\cdot10^{-14}$ & $7.7\cdot10^{-15}$\\
\noalign{\smallskip}\hline
\end{tabular}
\end{table}

\citet{bartlett_vanBuren_1986} gave the value of $S_{\text{Al},\text{Fe}}$ as $7\cdot10^{-13}$, and worsen it around five times to report a realistic limit of $4\cdot10^{-12}$, to reflect the limitations in the knowledge of the interior and the surface of the Moon, and to reflect the assumptions in their calculations. Taking the worse of the two values of $S_{\text{Al},\text{Fe}}$ mentioned in Table \ref{tab:res}, and using a scaling factor of five, our new limit on the violation of the equality of passive and active mass for Al and Fe gives $3.9\cdot10^{-14}$. If, however, the limit were taken from the k2d solution, it would, after using a scaling factor of five, give $3.4\cdot10^{-15}$.

Following \citet{clifford14}, based on the limit value of $S_{\text{Al},\text{Fe}}$, the limit value for $\zeta_3$ would also improve by about two orders of magnitude, assuming the same difference in binding energy between the atomic nuclei of Al and Fe. As mentioned in section \ref{sec:intro}, the value of $S_{\text{Al},\text{Fe}}$ is determined when differentiating between the crust and the mantle of the Moon. In reality, the lunar core will also add to the self force, and affect the value of $S_{\text{Al},\text{Fe}}$. For this study, to keep the assumptions the same as those used by \citet{bartlett_vanBuren_1986}, this effect was not considered. Furthermore, if considering a more recent value of the c.m.-c.f. offset from \citet{Smith_etal_2017}, the value of $S_{\text{Al},\text{Fe}}$ would become even smaller by a factor of 0.3, i.e. $2.5\cdot10^{-14}$. As mentioned earlier, such minor error sources are well captured by up-scaling the estimated error by a factor of five. 

\section{Conclusions}\label{sec:conclusions}
We determine a new limit on the violation of the equality of passive and active gravitational mass for Al and Fe following the procedure of \citet{bartlett_vanBuren_1986}. Our result benefits from the many years of very good \ac{LLR} data. We used different versions of our \ac{LLR} analysis software to obtain two final solutions, both providing the value of $\Delta\omega$/$\omega$. Using the model of \citet{bartlett_vanBuren_1986}, we convert the value of $\Delta\omega$/$\omega$ to obtain the coefficient $S_{\text{Al},\text{Fe}}$. If Al and Fe were to attract each other with the same force, the coefficient $S_{\text{Al},\text{Fe}}$ would be zero, otherwise this perturbation would affect the lunar orbit. We obtained two values for the limit of the ratio of active to passive mass for Al and Fe. Using the worse of the two values, we determine a new limit of $3.9\cdot10^{-14}$ on the possible violation of the equality of active and passive mass.

\section*{Acknowledgements}
Current LLR data are collected, archived, and distributed under the auspices of the International Laser Ranging Service (ILRS) \citep{ilrs2019}. We acknowledge with thanks that the processed LLR data, since 1969, has been obtained under the efforts of the personnel at the Observatoire de la C\^{o}te d'Azur in France, the LURE Observatory in Maui, Hawaii, the McDonald Observatory in Texas, the Apache Point Observatory in New Mexico, the Matera Laser Ranging observatory in Italy, and the Wettzell Laser Ranging System in Germany. This research was funded by the Deutsche Forschungsgemeinschaft (DFG, German Research Foundation) under Germany's Excellence Strategy EXC 2123 QuantumFrontiers, Project-ID 390837967. We also thank James G. Williams, California Institute of Technology for an extensive discussion on the effect of the degree 2 tides. Without his help, this publication would not have been possible.

\section*{Author contributions}
All authors contributed to the development of this study and provided ideas to its content. The first draft of the manuscript was written by VVS, and all authors commented on previous versions of the manuscript. All authors read and approved the final manuscript. 


\bibliography{references}




\acrodef{LLR}{Lunar Laser Ranging}
\acrodef{LUNAR}[LUNAR]{LUNar laser ranging Analysis softwaRe}
\acrodef{IERS}{International Earth Rotation and Reference Systems Service}
\acrodef{GMM}{Gauss-Markov model}

\end{document}